# Hexagonal-to-Cubic Phase Transformation in GaN Nanowires by Ga[+]-Implantation

Sandip Dhara,[#,$] Anindya Datta,[#,¥] Chi-We Hsu,[§,†] Chien-Ting Wu,[#,] Ching-Hsing Shen,[§] Zon-Huang Lan,[#] Kuei-Hsien Chen,[#] Li-Chyong Chen,[§,*] Yuh-Lin Wang,[#,] Chia-Chun Chen[†]

[#]Institute of Atomic and Molecular Sciences, Academic Sinica, Taipei 106, Taiwan

[§]Center for Condensed Matter Sciences, National Taiwan University, Taipei 106, Taiwan

[†]Department of Chemistry, National Taiwan Normal University, Taipei 116, Taiwan

Department of Physics, National Taiwan University, Taipei 106, Taiwan

Abstract

Hexagonal to cubic phase transformation is studied in focused ion beam assisted Ga+-implanted GaN nanowires. Optical photoluminescence and cathodoluminescence studies along with high-resolution transmission electron microscopic structural studies are performed to confirm the phase transformation. In one possibility, sufficient accumulation of Ga from the implanted source might have reduced the surface energy and simultaneously stabilized the cubic phase. Other potential reason may be that the fluctuations in the short-range order induced by enhanced dynamic annealing (defect annihilation) with the irradiation process stabilize the cubic phase and cause the phase transformation.

* Corresponding author E-mail: chenlc@ccms.ntu.edu.tw

[$] On leave from Materials Science Division, Indira Gandhi Centre for Atomic Research, Kalpakkam-603102, India

[¥] Now at Netaji Nagar Day College, 170/436 N.S.C Bose Road, Kolkata-700092, India

Interest in the growth and physical properties of GaN originates from its potential applications as near-UV optoelectronic, and high temperature, high speed devices.[1-3] High ionicity in III-V compounds makes GaN thermally stable only in the hexagonal wurtzite (h-GaN) structure.[4] Metastable cubic zinc-blend (c-GaN) modification of this material is extremely difficult in equilibrium condition. However, growth of epitaxial (epi-) c-GaN layer has been achieved on different substrates (Si, 3C-SiC, MgO, GaP and GaAs) mostly using the nonequilibrium growth conditions of plasma assisted molecular beam epitaxy (MBE)[2,5] and low pressure metalorganic chemical vapor deposition (LP-MOCVD)[6] techniques. Cubic phase has potential superiority over hexagonal phase for certain device applications and integrity of GaN devices on readily available substrates.[7] For example, c-GaN is expected to have a higher mobility, resulting from its lower phonon scattering in the higher crystallographic symmetry.[8] Possibilities have already been demonstrated for getting high p-type conductivity in c-GaN.[9] It is also predicted theoretically that the optical gain in c-GaN quantum wells might be higher than that in hexagonal GaN wells.[10] The successful production of high quality h-GaN nanowires (NWs)[11] using vapor-liquid-solid (VLS) technique has opened up the possibility of device application in these one-dimensional (1-D) system. As a matter of fact, single nanorod h-GaN laser has already been demonstrated.[12] However, formation of pure c-GaN NW is not so far demonstrated to utilize the various advantages over the hexagonal phase.

In case of epi-film the metastable c-GaN is formed by providing lattice matching substrates for epitaxial growth and by kinetic (nonequilibrium processing using techniques like plasma assisted MBE[2,5] or LPCVD[6]) or thermodynamic stabilization ('shift of equilibrium' adopting either Ga-rich[5-9] conditions or adding surfactant[13] like As to lower the surface energy [14]) of the metastable phase. For the growth of nanostructures in the VLS process,[15] however, none of these techniques can be effectively adopted to stabilize the metastable cubic phase in a controlled fashion. Instead, one can think of nonequilibrium modification of the nanostructured materials to transform the stable phase to a metastable one using energetic irradiation process. Stabilization of metastable phases (order-disorder or coherent phase transition preserving the underlying lattice structure of the stable phase) using energetic ion irradiation is well reported in alloy systems.[16] Metastable c-BN is also reported to nucleate with medium/high energy electron or low energy ion irradiation (shallow subsurface implantation; 'subplantation' technique)[17] on h-BN film at room temperature (RT).[18] Most interestingly, phase transformation is observed for graphitic carbon nano-onions to diamond by high energy electron irradiation at slightly elevated temperature (675 K) in absence of pressure or catalyst.[19] In last two cases, the phase transformation falls in the category of incoherent phase transition with two phases containing same element(s) but different crystal symmetry.[17-19] Nano-diamond is also reported to form by ion (mixture of Ar, $H_2$ and $CH_4$) beam irradiation on amorphous carbon film.[20] Ion beam induced stress and possible role of atomic hydrogen/hydrogen ions in stabilizing the $sp^3$ bonds are discussed as the origin of the phase transformation.

We report here a phase transformation of GaN NWs from stable wurtzite to metastable zinc-blend structure induced by focused ion-beam (FIB) assisted $Ga^+$-ion implantation. Optical photoluminescence (PL) and cathodoluminescence (CL) studies along with high-resolution transmission electron microscopic (HRTEM) structural studies are performed to confirm the phase transformation.

Randomly oriented GaN NWs were grown by chemical-vapor deposition technique following VLS process. The samples were grown at $900^0$C on c-Si substrates pre-coated with Au catalyst, using molten gallium as source material and $NH_3$ (10 sccm) as reactant gas in a horizontal tubular furnace. $Ga^+$ implantation on these GaN NWs was achieved using a FIB at 50 keV in the fluence range of $1 \times 10^{14}$ - $2 \times 10^{16}$ ions $cm^{-2}$ with an ion flux of ~ $5 \times 10^{12}$ ions $cm^{-2}$ $sec^{-1}$. As calculated from the SRIM code,[21] the range of 50 keV $Ga^+$ in GaN film is 24 nm with energy dissipation mainly through nuclear energy loss. Details of ion irradiation process is reported elsewhere.[22] PL measurements were performed using He-Cd laser tuned to 325 nm with an output power of ~10 mW at room temperature. The emission signal was collected by a SPEX 0.85-m double spectrometer and detected by a lock-in-amplifier. Temperature dependent CL study was performed using a Gatan MonoCL3 system attached to a Jeol JSM-6700F field emission scanning electron microscope (FESEM). Structural studies were performed using JEOL JEM-4000EX high-resolution transmission electron microscopy (HRTEM).

FESEM image (Figure 1a) showed randomly oriented pristine NWs. HRTEM image (Fig. 1b) showed formation of wurtzite GaN phase with zone axis lying along [001] direction as calculated from the corresponding diffraction pattern (inset Figure 1b). The growth of the typical NW was observed to be along [100] direction.

Room temperature PL study of the pristine NWs showed two peaks in the range of ~ 3 - 4 eV (Figure 2a). A small peak around 3.1 eV might correspond to longitudinal optical (LO) phonon replica mode of the donor-acceptor pair (DAP) for cubic phase.[11, 23] The minor cubic phase could be stable as stacking faults embedded in the h-GaN NW, as reported in our earlier study.[11] Intense peak at 3.42 eV closely corresponded to direct band-to-band transition ($E_g$) energy of h-GaN[24] confirming presence of predominant wurtzite phase in the pristine NWs. After FIB irradiation with 50 keV $Ga^+$ we observed continuous deterioration (Figure 2a) of intensities for both the peaks with increasing fluence. This might be due to the increase in disorder introduced during the irradiation process with increasing fluence. Most interestingly, intensity of the peak at 3.42 eV corresponding to the major wurtzite phase was observed to deteriorate faster than that of the peak around 3.1 eV correlated to minor phonon replica mode for cubic phase. This gave rise to an increasing peak intensity ratio ($I_C/I_H$) for minor cubic to major hexagonal phase with increasing fluence (inset Figure 2a), indicating a possible hexagonal-to-cubic phase transformation in the irradiated NWs. We did not show the data corresponding to the NWs irradiated at $2\times10^{16}$ ins $cm^{-2}$ as complete amorphization took place at this fluence, reported in our earlier study dealing with the defect evolution of the as-irradiated NWs.[22]

In a next step, we annealed the as-irradiated NWs by a two-step process, primarily to remove the defects and finally to examine the realistic nature of the phase transformation in well crystalline NWs. The two-step annealing process was performed with 15 minutes at 650 $^0$C and 2 minutes at 1000 $^0$C in $N_2$ ambient, as recommended in the literature for the irradiated epi-GaN.[25] With the reappearance of the peak around 3.42 eV, room temperature PL study for the post-annealed NWs showed presence of both the peaks with reduced intensity as compared to that for the pristine NWs (Figure 2b). The positions of the peak corresponding to $E_g$ were shifted towards lower energy ('redshift'). The 'redshift' of the peak position corresponding to $E_g$ in the post-annealed NWs was appreciable for the sample irradiated above a fluence of $1\times10^{15}$ ions cm$^{-2}$ as indicated from the preliminary study of the hexagonal-to-cubic phase transformation in the as-irradiated NWs (inset Figure 2a). However, the 'redshift' was reduced in the post-annealed sample irradiated at a high fluence of $1\times10^{16}$ ions cm$^{-2}$. The 'redshift' was maximum for the post-annealed NWs irradiated at an optimum fluence of $5\times10^{15}$ ions cm$^{-2}$. The peak was shifted to 3.28 eV, which is close to the direct band-to-band transition peak reported for cubic (zinc-blend) phase at RT.[23] At the outset, it should be mentioned that upon annealing further crystallization was also not observed for the NWs irradiated at fluence of $2\times10^{16}$ ions cm$^{-2}$ showing complete amorphization[22] in the as-irradiated sample.

Temperature dependent CL study of the post-annealed NWs irradiated at an optimum fluence of $5\times10^{15}$ ions cm$^{-2}$ showed (Figure 3a) an expected 'blueshift' with decreasing temperature. $E_g$ peaked around 3.25 eV and 3.3 eV in RT and 4K, respectively, confirming the presence of cubic phase in the system.[26] Small difference in the band-to-band transition peak energy in room temperature PL and CL measurements fell within the limit of discrepancy reported for the GaN system.[27] The fitting (inset Figure 3a) of the temperature dependence of peak energies of the band-to-band transition matched well with the reported values of Varshni parameters [$\alpha = 6.697\times10^{-4}$, Deby Temperature ($\Theta$) of GaN = 600K; $E_g(T) = E_0 - \alpha T^2/(T+\Theta)$ where calculated $E_g$ for cubic phase at 0K ($E_0$) = 3.302 eV] for c-GaN.[26] The defect band around 3.1 eV corresponding to LO phonon replica mode of cubic phase observed in the room temperature PL studies could not be traced at low temperature CL plots. The asymmetry in the RT-CL plot, however indicated the presence of peak around 3.1 eV. The comparison of secondary electron (SE) and CL (excited at 3.3 eV) images (Figure 3b) at 4K shows large of the NWs are in cubic phase.

HRTEM analysis of the post-annealed NWs always showed h-GaN as major phase with some amount of cubic phase embedded (not shown in picture) in the NWs irradiated at $2\times10^{15}$ ions cm$^{-2}$. The post-annealed NWs irradiated at an optimum fluence of $5\times10^{15}$ ions cm$^{-2}$ showed (Figure 4a) the presence of single cubic phase with zone axis lying along [0 –1 1] direction and lattice parameter was calculated to be 0.455 nm, close to the reported value for c-GaN.[5] The lattice parameter in cubic phase was increased to 0.47 nm for the post-annealed NWs irradiated at a fluence of $1\times10^{16}$ ions cm$^{-2}$ with zone axis lying along [0 0 1] direction (Figure 4b). The increase in the lattice parameter showed presence of strain in the post-annealed NWs irradiated above the optimum fluence causing shift of peak energy corresponding to $E_g$ towards higher energy than that observed for the NWs irradiated at an optimum fluence (Figure 2b).

The phase transformation observed at an optimum fluence of $5\times10^{15}$ ions cm$^{-2}$ may be due to sufficient accumulation of Ga from the implanted source in reducing the surface energy and simultaneously stabilizing the cubic phase which is well documented in the literature.[13,14] However, we can further explore the role of defects created in the irradiation process to understand the phase transition in details. In our earlier study,[22] we have reported enhanced dynamic annealing in as-irradiated GaN NWs with short-range order appearing just below an optimum fluence of $5\times10^{15}$ ions cm$^{-2}$ for efficient annealing. Ion-beam-generated Frenkel pairs, which survive the ultra short duration quenching ($\sim10^{-10}$ sec) of collision cascades, are mobile in h-GaN and a vacancy (or interstitial) cluster is likely to experience annihilation by trapping of mobile interstitials. High diffusivity of mobile point-defects (interstitials) in the high-curvature geometry of NWs enhances the dynamic annealing effect introducing short-range order in as-irradiated NWs around the optimum fluence. The most likely fluctuations that can lead to nucleation sites for the second phase are those that are short ranged. This is an experimentally observed fact, particularly, in case of first-order phase transition where the phenomenon occurs discontinuously.[28] The phase transition takes place in an inhomogeneous nucleation process where the nucleation starts about particular sites and grows to encompass the whole sample. The nucleated second phase at the defect sites may dominate during recrystallization upon annealing.

Formation of large amount of nitrogen vacancy ($V_N$) and Ga interstitial ($Ga_I$) (formation energy of $V_N \approx 4$ eV/atom and $Ga_I \approx 10$ eV/atom in GaN)[29] are expected in this nuclear energy-loss dominated process with 50 keV Ga$^+$ irradiation on GaN NWs. Presence of these defects in as-irradiated NWs deteriorates the crystalline quality making the band-to-band transition peak invisible in our PL study (Figure 2a). Two-step annealing process in N$_2$ ambient gets back the crystalline quality, at least partially, to make the band-to-band transition peak observable in the PL study of the post-implanted samples (Figure 2b). The post-implanted NWs irradiated with a fluence upto $1\times10^{15}$ ions cm$^{-2}$ do not show appreciable change in the peak position corresponding to $E_g$ with respect to that for the h-GaN. This may be due to insufficient nucleation for cubic phase upto this fluence. We have started observing the presence of cubic phase for the post-annealed NWs irradiated with fluences above $1\times10^{15}$ ions cm$^{-2}$. Distinct 'redshift' is observed in the peak position corresponding to $E_g$ for the post-annealed NWs irradiated at a fluence of $2\times10^{15}$ ions cm$^{-2}$ (Figure 2b). This may be due to the presence of minor cubic phase, which may have nucleated at the defect sites with the short-range order appearing in as-irradiated NWs around this fluence. 'Redshift' of peak corresponding to $E_g$ is also reported for GaN sample with mixed cubic and hexagonal phases.[27,30] Near stoichiometric cubic phase is observed (Figure 4a) for majority of the post-annealed NWs irradiated at an optimum fluence of $5\times10^{15}$ ions cm$^{-2}$ indicating sufficient accumulation of nucleation sites for the cubic phase in as-irradiated NWs. The 'redshift' for the peak position corresponding to $E_g$ is also maximum and the peak position is close to that reported for the cubic phase. The CL image, taken with excitation energy 3.3 eV at 4K, of the post-annealed NWs irradiated at an optimum fluence shows large numbers of NWs in cubic phase (Figure 3b). With dynamic annealing being inefficient beyond the optimum fluence,[22] the defects in the irradiated NWs become too high to be removed by similar annealing treatment. Thus, the 'redshift' occurred for the peak position corresponding to $E_g$ is small (Figure 2b) for the post-annealed NWs irradiated at a high fluence of $1\times10^{16}$ ions cm$^{-2}$ even with sufficient nucleation sites for cubic phase. Structural study showed the nucleation of cubic phase but with strained lattice (Figure 4b) causing the shift of the peak energy corresponding to $E_g$ to a slightly higher side than that observed for c-GaN for the post-annealed NWs irradiated at an optimum fluence of $5\times10^{15}$ ions cm$^{-2}$. Thus, at an optimized fluence with sufficient accumulation of nucleation sites for the cubic phase a near-complete phase transition can occur between hexagonal-to-cubic phases of GaN, which is incoherent in nature involving same elements but two different crystal symmetries in the system.

In conclusion, a hexagonal-to-cubic phase transformation, as suggested by luminescence study along with direct evidence supported by transmission electron microscopy, in GaN NWs is observed with $Ga^+$-implantation above an optimum fluence of $5x10^{15}$ ions $cm^{-2}$. The exact origin, either Ga playing a role of surfactant reducing the surface energy or the fluctuations in the short-range order induced by enhanced dynamic annealing with the irradiation process, in stabilizing the cubic phase and causing the phase transformation to occur can not be determined definitely. This requires further study. Nevertheless, the formation of c-GaN NWs, with its various technological advantages over h-GaN, will be useful for futuristic device applications.

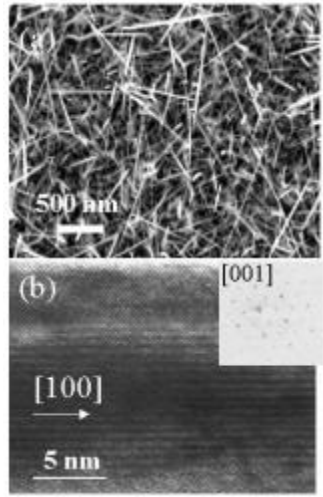

Figure 1. (a) FESEM image showing randomly oriented pristine NWs. (b) HRTEM lattice image of the pristine GaN NW and corresponding selected area electron diffraction in the inset, showing formation of wurtzite GaN phase with zone axis lying along [001] direction. The growth direction of the typical NW is along [100] direction.

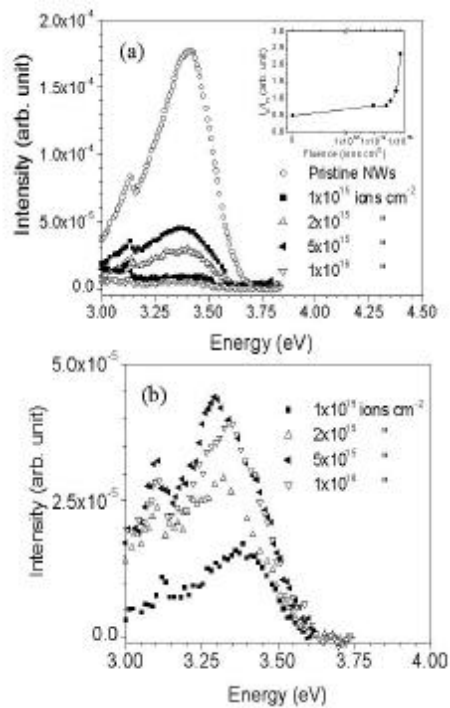

Figure 2. (a) Room temperature PL spectra for the pristine and as-irradiated NWs showing rapid deterioration of peak corresponding to direct band-to-band transition energy with increasing fluence. Inset shows fluence dependence of peak intensity ratio for the cubic (LO phonon replica mode of the DAP at ~ 3.1 eV) and hexagonal (direct band-to-band transition energy at 3.42 eV) phases. (b) Room temperature PL spectra for the post-annealed NWs irradiated with different fluences showing reappearance of peak corresponding to direct band-to-band transition energy with a 'redshift' above a fluence of $1\times10^{15}$ ions cm$^{-2}$.

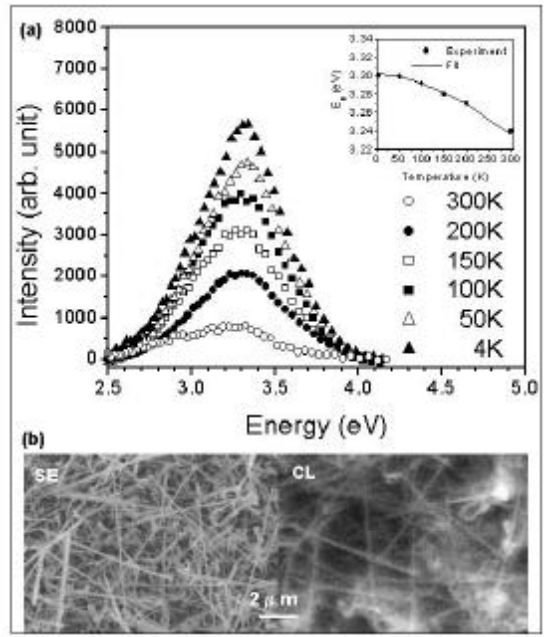

Figure 3. (a) Temperature dependent CL spectra for the post-annealed NWs irradiated with an optimum fluence of $5\times10^{15}$ ions cm$^{-2}$ showing presence of cubic GaN. Inset shows the temperature dependence of the band-to-band transition peak energies. (b) Secondary electron (SE) and CL (excited at 3.3 eV) images at 4K showing large of the NWs are in cubic phase.

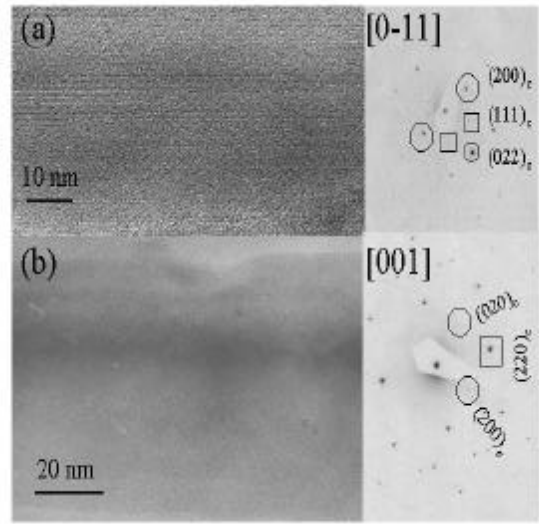

Figure 4. (a) HRTEM lattice image of the post-annealed NW irradiated at an optimum fluence of $5\times10^{15}$ ions cm$^{-2}$ and corresponding selected area electron diffraction in the inset, showing formation of cubic GaN phase with zone axis lying along [0-11] direction. (b) HRTEM lattice image of the post-annealed NW irradiated at a fluence of $1\times10^{16}$ ions cm$^{-2}$ and corresponding selected area electron diffraction in the inset, showing formation of cubic GaN phase with zone axis lying along [001] direction.

12